\documentclass[aps,prl,twocolumn,superscriptaddress, footinbib]{revtex4}
\pdfoutput=1
\usepackage[utf8]{inputenc}
\usepackage{mathtools}
\usepackage{amsmath}
\usepackage{amssymb}
\usepackage{esint}
\usepackage{mathtools}
\usepackage{bm} 
\usepackage{natbib}

\newcommand{\prlsec}[1]{\textit{#1.---}}
\renewcommand{\vec}[1]{\bm{#1}}

\let\oldalpha\alpha
\renewcommand{\alpha}{\bm{\oldalpha}}
\let\oldbeta\beta
\renewcommand{\beta}{\bm{\oldbeta}}

\let\oldhat\hat
\renewcommand{\hat}[1]{\oldhat{\mathbf{#1}}}

\let\olddelta\delta
\renewcommand{\delta}{\olddelta^{(3)}}

\begin{document}

\title{General Galilei Covariant Gaussian Maps}

\author{Giulio Gasbarri}

\email{giulio.gasbarri@ts.infn.it}

\affiliation{Abdus Salam ICTP, Strada Costiera 11, I-34151 Trieste, Italy} 

\affiliation{Department of Physics, University of Trieste, Strada Costiera 11,
34151 Trieste, Italy}

\affiliation{Istituto Nazionale di Fisica Nucleare, Trieste Section, Via Valerio
2, 34127 Trieste, Italy}

\author{Marko Toro\v{s}}

\email{marko.toros@ts.infn.it}

\affiliation{Department of Physics, University of Trieste, Strada Costiera 11,
34151 Trieste, Italy}

\affiliation{Istituto Nazionale di Fisica Nucleare, Trieste Section, Via Valerio
2, 34127 Trieste, Italy}

\author{Angelo Bassi}

\email{bassi@ts.infn.it}

\affiliation{Department of Physics, University of Trieste, Strada Costiera 11,
34151 Trieste, Italy}

\affiliation{Istituto Nazionale di Fisica Nucleare, Trieste Section, Via Valerio
2, 34127 Trieste, Italy}

\begin{abstract}
We characterize general non-Markovian Gaussian maps which are covariant under Galilean transformations. In particular, we consider translational and Galilean covariant maps and show that they reduce to the known Holevo result in the Markovian limit. We apply the results to discuss measures of macroscopicity based on classicalization maps, specifically addressing dissipation, Galilean covariance and non-Markovianity. We further suggest a possible generalization of the macroscopicity measure defined in Nimmrichter and Hornberger [Phys. Rev. Lett. 110, 16 (2013)].  
\end{abstract}

\maketitle

\prlsec{Introduction} Symmetries have always played a central role in modern physics, especially after their mathematical formulation with the advent of group theory: they underlie the simplicity of nature and manifest the beauty of physical laws. They also serve as a guideline principle for deciding the form of the dynamics \cite{feynman1966feynman,peskin1996introduction}. Here we are interested in the role of space-time symmetries in non-relativistic quantum mechanics. 

The covariance of the Schr\"{o}dinger equation, and of the corresponding Liouville-von Neumann equation, under the action of the Galielan group, has been extensively discussed~\cite{bargmann1954unitary,ballentine1998quantum,pereira2015galilei}.
On the other hand, the investigation of Galilean covariance within the context of open quantum systems is still an area of active research~\cite{breuer2002theory}.
The exact quantum dynamics of a system interacting with the surrounding environment can be very complicated: in general, heavy approximations and heuristical arguments are needed in order to arrive at an explicit useful expression for the system's effective dynamics. In this case, symmetries can be a guiding principle in constructing the effective dynamics, bypassing at least partially the complexity (or impossibility) of a direct calculation by imposing constraints, which are expected to hold not only at the fundamental level, but also at the effective level~\cite{vacchini2001translation, de2013derivation, yan2000unified, yan2000unified1, Sandulescu1987277,474887, vacchini2009quantum, 6b00498,ferialdi2016exact}.

Space-time symmetries in open quantum systems have been fully analyzed only in the special, but very important, case of a Markovian, completely positive (CP) and trace preserving (TP) dynamics. This dynamics, discussed in the seminal works of Gorini, Kossakowski, Sudarshan and independently by Lindblad \cite{gorini1976completely,lindblad1976generators}, is known as the \emph{quantum dynamical semigroup}: it is generated by the \emph{Lindblad superoperator} and can be written as a first order differential equation, called the \emph{Lindblad master equation}.  
By imposing the additional request of covariance under the action of the Galilei group, Holevo in a series of works \cite{holevo1993note,holevo1993conservativity,
holevo1995translation,holevo1996covariant}, completely \emph{characterized} translational and Galilei covariant Lindblad master equations, by giving the explicit form of the Lindblad superoperators \footnote{The characterization is unique up to a unitary transformation (the same applies also for non-Markovian Gaussian maps discussed in this text).}.

The Holevo characterizations play a major role in the description of several important physical phenomena such as environmental decoherence and relaxation phenomena~\cite{vacchini2001translation, de2013derivation, yan2000unified, yan2000unified1, Sandulescu1987277,474887, vacchini2009quantum, 6b00498}. Furthermore, they are also relevant for the foundation of quantum mechanics, where an intrinsic non-unitary dynamics is postulated to solve the measurement problem \cite{bassi2003dynamical, bassi2013models, weinberg2016happens}, the black hole information paradox~\cite{hawking1976breakdown}, or to combine principles of general relativity with quantum mechanics~\cite{penrose1996gravity}.

Although the assumption of Markovianity is often well justified, recent technological advances have lead to investigating several phenomena exhibiting memory effects \cite{breuer2016colloquium}, e.g., ultrafast chemical reactions~\cite{plotkin1998non,burshtein2004non,pomyalov2005non, gindensperger2006short,guerin2012non,chiarugi2015modelling}, side band cooling~\cite{triana2016ultrafast} and light harvesting in photosynthesis~\cite{pachon2012computational,pachon2011physical,schroder2006calculation,plenio2008dephasing,thorwart2009enhanced,nalbach2011exciton}. 
This is of little surprise, as the \emph{time resolution} of experimental apparatuses has increased severalfold in the last decades.
It is therefore now clear that non-Markovian dynamics will acquire a more prominent role in the near future: the theoretical investigations are pressed by  practical necessity. 

In this letter we will derive the general structure of non-Markovian Galilei covariant Gaussian maps. More specifically, we will consider the non-Markovian Gaussian map introduced in Ref.~\cite{diosi2014general}, and we will impose covariance under Galilean space-time symmetries (translations, boosts and rotations). In this way we will obtain a generalization of the Holevo generators \cite{holevo1993note,holevo1993conservativity,
holevo1995translation,holevo1996covariant} to the non-Markovian Gaussian case. Using these results, we will discuss measures of macroscopicity based on classicalization maps. Specifically, we will address the role of non-Markovian and dissipative effects, which limit the validity of the macroscopicity measure proposed in Ref.~\cite{nimmrichter2013macroscopicity}.

\prlsec{General framework of Gaussian maps} Non-Markovian dynamics are in general difficult to analyze: the system and environment form a complicated many-body problem which, without some additional simplifying assumption, remains intractable. On the other hand, the subclass of (non-Markovian) Gaussian maps, still appropriate for the description of a vast spectrum of phenomena~\cite{tavis1968exact, leggett1987dynamics, hu1992quantum, diosi1998non, greentree2006, adler2007collapse, pachon2010nonclassical, diosi2012non, guo2012critical, pachon2013mechanisms, wang2015nonequilibrium, estrada2015quantum,tilloy2017interacting}, can be analyzed both analytically~\cite{diosi2014general} and numerically~\cite{stockburger2002exact, tilloy2017markovian}. 

The starting point of our analysis is the most general trace-preserving, completely positive Gaussian map derived in Ref.~\cite{diosi2014general} (we work in interaction picture and adopt Einstein's
summation convention):
\begin{align}\label{eq:df}
\mathcal{M}_{t}=&\exp_{+}\left\{ \int_{0}^{t}d\tau\int_{0}^{t}dsD_{jk}(\tau,s)\times \right. \nonumber \\
& \times \left.\left(\oldhat{A}_{sL}^{k}\oldhat{A}_{\tau R}^{j}-\theta_{\tau s}\oldhat{A}_{\tau L}^{j}\oldhat{A}_{sL}^{k}-\theta_{s \tau}\oldhat{A}_{sR}^{k}\oldhat{A}_{\tau R}^{j}\right)\right\}, 
\end{align}
where $\exp_{+}$ denotes the time-ordered exponential, $D_{jk}$ is a complex valued positive semi-definite matrix, $\oldhat{A}$ are bounded Hermitian operators and the subscript $L$ ($R$) denotes operators acting on the statistical operator ${\rho}$
from the left (right), e.g. $\oldhat{A}_{L}^{k}\oldhat{A}_{R}^{j}\oldhat{\rho}=\oldhat{A}^{k}\oldhat{\rho}\oldhat{A}^{j}$ with $\oldhat{A}^{k}$ Hermitian operators. The correlation matrix $D_{jk}(\tau,s)$ and the operators $\oldhat{A}^{k}$ are supposed to encode, phenomenologically, the action of the bath on the system. We note that, by imposing the request of Markovianity:
\begin{equation}
D_{jk} (\tau, s)=\olddelta(\tau-s)\tilde{D}_{jk}(s),
\label{eq:markovianity}
\end{equation}
where $\tilde{D}_{jk}(s)$ is a complex valued positive semi-definite matrix, the exponent in Eq.~\eqref{eq:df} takes the well-known Lindblad form. 

Since we are interested in space-time symmetries, we now explicitly assume that the Hilbert space $\mathcal{H}_{\mathcal{S}}$ is $L^{2}(\mathbb{R}^{3})$ (the generalization to the $N$-particle Hilbert space is straightforward).
In this case it is convenient to decompose the operators in Eq.~\eqref{eq:df}
by using the Weyl-Wigner decomposition (in Schr\"odinger picture) \cite{weyl1927quantenmechanik}: 
\begin{align}
\hat{A}_t= & \int_{\mathbb{R}^3} d\alpha \int_{\mathbb{R}^3} d\beta\, \mathcal{A}_{t}(\alpha,\beta)e^{{i}(\alpha\cdot\hat{x}+\beta\cdot\hat{p})},\label{eq:decomGM}
\end{align}
where $\hat{A}_t$ may depend explicitly on time, which is encoded in the time-dependency of $\mathcal{A}_{t}$, and  $\hat{x}$ and $\hat{p}$ are the standard position and momentum operators. It is then straightforward to show that the map in Eq.~\eqref{eq:df} becomes (in the interaction picture):
\begin{align}
\mathcal{M}{}_{t}=\exp_{+}\bigg\{&\int\!\!dT\!\!\int\!d\Gamma\,\mathcal{D}(\alpha_{1},\beta_{1},\alpha_{2},\beta_{2},\tau,s) 
\Theta^{\mu \nu}_{\tau s} \nonumber\\
\times\Big(&e^{i(\alpha_{1}\cdot\hat{x}_{s\mu}+\beta_{1}\cdot\hat{p}_{\mu})}e^{-i(\alpha_{2}\cdot\hat{x}_{\tau \nu}+\beta_{2}\cdot\hat{p}_{\nu})}\Big)\bigg\},
\label{eq:rho_formal2}
\end{align}
where  $d T= d\tau ds$, $d\Gamma= d\alpha_{1}d\beta_{1}d\alpha_{2}d\beta_{2}$, the integration domains, which we omit to simplify the notation, are $[0,t]\times[0,t]$  and $\otimes_{j=1}^4\mathbb{R}^{3}$ for the $T$ and $\Gamma$ integrals, respectively, $\hat{x}_{s}$ is the position operator in the interaction picture at time $s$, $\mu$ and $\nu$ denote $L$ or $R$ (left or right operators), $\Theta_{\tau s}^{LR}=\Theta_{\tau s}^{RL}=1/2$, $\Theta_{\tau s}^{LL}=-\theta_{\tau s}$, $\Theta_{\tau s}^{RR}=-\theta_{s \tau}$ and
\begin{equation}
\mathcal{D}(\alpha_{1},\beta_{1},\alpha_{2},\beta_{2},\tau,s)=D_{jk}(\tau,s)\mathcal{A}_{\tau}^{j\,*}(\alpha_{1},\beta_{1})\mathcal{A}_{s}^{k}(\alpha_{2},\beta_{2})\label{eq:kernel}
\end{equation}
is a kernel that satisfies the following symmetry property ~\footnote{From the definition of $D_{j k}(\tau,s)$  in \cite{diosi2014general} we have $D_{k j}^*(s,\tau)=D_{j k}(\tau,s)$.}: 
\begin{align}
\mathcal{D}(\alpha_{1},\beta_{1},\alpha_{2},\beta_{2},\tau,s)=\mathcal{D}^{*}(\alpha_{2},\beta_{2},\alpha_{1},\beta_{1},s,\tau).
\end{align}
We now impose the relevant Galilei symmetry on the system, constraining the form of the dynamics given by Eq.~\eqref{eq:rho_formal2}.

\prlsec{Covariance} Let us consider a locally compact Lie group $G$ and a unitary representation $\hat{U}_{g}$, with $g \in G$, on the Hilbert space of the system. Following~\cite{davies1970d,holevo1973statistical} a quantum dynamical map is said to be G-covariant if it commutes with the linear transformation $\mathcal{U}_{g}[\,\cdot\,]=\hat{U}_{g}\,\cdot\,\hat{U}_{g} $:
\begin{align}\label{eq:semi}
\mathcal{M}_{t}= \mathcal{U}_{g}^{-1}\circ\mathcal{M}_{t}\circ\mathcal{U}_{g}.
\end{align}

With reference to the single particle Hilbert space $\mathcal{\ensuremath{H}}_{\mathcal{S}}$ $(L^{2}(\mathbb{R}^{3}))$ we assume that  the Hamiltonian is covariant under the relevant symmetry of the Galilei group $\mathcal{G}$~\footnote{In the one particle case considered here it means that the Hamiltonian is $\hat{H}=\hat{p}^{2}/2m$. The many particle case, which is physically richer, as particle-particle interactions are allowed, can be treated in a similar way. See the discussion in~\cite{bargmann1954unitary,ballentine1998quantum,pereira2015galilei} for a general discussion of the covariance of the dynamics (of isolated systems), as well as~\cite{foldy1961relativistic}, for the covariance of the center of mass dynamics.}: specifically, we consider the centrally extended unitary representation ($\hat{U}_g$) of the Galilei group ($\mathcal{G}$) on $\mathcal{\ensuremath{H}}_{\mathcal{S}}$. The generators of infinitesimal translations, boosts and rotations are (in the interaction picture): 
\begin{align}
\hat{P}=&\, \hat{p},\label{eq:algebraP}\\
\hat{J}=&\, \hat{x}\times\hat{p},\label{eq:algebraJ}\\
\hat{K}=&\, m\hat{x},\label{eq:algebraK}
\end{align}
respectively, where $m$ is the mass of the particle. Exploiting Eq.~\eqref{eq:rho_formal2}, and the fact that we are considering
a unitary representation, it is straightforward to show that Eq.~\eqref{eq:semi} is satisfied if and only if 
the following condition is satisfied: 
\begin{align}
 &\int\!\!dT\!\!\int\!d\Gamma\,\mathcal{D}(\alpha_{1},\beta_{1},\alpha_{2},\beta_{2},\tau,s)\Theta^{\mu \nu}_{\tau s} \nonumber \\
 &\hspace{1.3cm}\Big(e^{i(\alpha_{1}\mathcal{U}_{g}[\hat{x}_{s \mu}]+\beta_{1}\mathcal{U}_{g}[\hat{p}_{\mu}])}e^{-i(\alpha_{2}\mathcal{U}_{g}[\hat{x}_{\tau \nu}]+\beta_{2}\mathcal{U}_{g}[\hat{p}_{\nu}])}\nonumber\\
&\hspace{1.3cm}-e^{i(\alpha_{1}\hat{x}_{s \mu}+\beta_{1}\hat{p}_{\mu})}e^{-i(\alpha_{2}\hat{x}_{\tau \nu}+\beta_{2}\hat{p}_{\nu})}\Big)=0.\label{eq:1cond}
\end{align}
This equation constrains the structure of the dynamical map under the Galilean symmetry $g\in \mathcal{G}$. In particular, we will now see how the request of translation (boost) covariance characterizes the structure of the dynamical map.

\prlsec{Translational-covariance}
Restricting to the subgroup of translations $ \mathcal{T}\subset \mathcal{G}$ we have that: 
\begin{align}
\mathcal{U}_{\vec{a}}[\hat{x}_{t}] & =\hat{x}_{t}+\vec{a},\label{eq:tx}\\
\mathcal{U}_{\vec{a}}[\hat{p}] & =\hat{p},\label{eq:tp}
\end{align}
where $\hat{x}_{t}=\hat{x}+(\hat{p}/{m})t$ is the position operator in the interaction picture at time $t$, $\vec{a}$ is a translation vector and $\mathcal{U}_{\vec{a}}$ denotes the corresponding linear transformation (see Eq.~\eqref{eq:semi}). Using Eqs. (\ref{eq:tx}), (\ref{eq:tp})
we obtain from Eq.~\eqref{eq:1cond}: 
\begin{align}
&\int \!\!dT\!\!\int\!d\Gamma\,\mathcal{D}(\alpha_{1},\beta_{1},\alpha_{2},\beta_{2},\tau,s) \Theta^{\mu \nu}_{\tau s}\nonumber\\
&e^{i(\alpha_{1}\cdot\hat{x}_{\mu s}+\beta_{1}\cdot\hat{p}_{\mu})}e^{-i(\alpha_{2}\cdot\hat{x}_{\nu \tau}+\beta_{2}\cdot\hat{p}_{\nu})}(1-e^{i(\alpha_{1}-\alpha_{2})\cdot \vec{a}})=0.
\label{eq:condbis}
\end{align}
Since this relation must hold $\forall \vec{a}$, it follows that
Eq.~\eqref{eq:condbis} is satisfied if and only if the following equality holds 
\begin{align}
\mathcal{D}(\alpha_{1},\alpha_{2},\beta_{1},\beta_{2},\tau,s)=
&\delta(\alpha_{1}-\alpha_{2}) \nonumber\\
&\times\mathcal{D}_{T}(\alpha_{1},\alpha_{2},\beta_{1},\beta_{2},\tau,s),\label{eq:condexpt}
\end{align}
where $\mathcal{D}_{T}$ is a complex valued function, which we rewrite as:
\begin{align}
 \mathcal{D}_{T}(\alpha_{1},\beta_{1},\alpha_{2},\beta_{2},\tau,s)\!=& D_{jk}(\tau,s)\nonumber \\
&\tilde{\mathcal{A}}_{\tau}^{j*}(\alpha_{1},\beta_{1})
\tilde{\mathcal{A}}_{s}^{k}(\alpha_{2},\beta_{2}).
\label{eq:D_T}
\end{align}

We then insert Eq.~\eqref{eq:condexpt} into Eq.~\eqref{eq:rho_formal2}, use Eq.~\eqref{eq:D_T}, integrate over $\alpha_2$ and relabel  $\alpha_1$  as $\alpha$ to obtain:
\begin{align}
\mathcal{M}_{t}= & \exp_{+}\bigg\{\int_{0}^{t}\!\!d\tau \int_{0}^{t} d s \!\!\int_{\mathbb{R}^3}\!d\alpha\,D_{jk}(\tau,s)\nonumber\\
&\Big([F_{s L}^{k}(\hat{p},\alpha)e^{i\alpha\cdot\hat{x}_{s L}}]\,[F_{\tau R}^{j\dagger}(\hat{p},\alpha)e^{-i\alpha\cdot\hat{x}_{\tau R}}]\nonumber\\
&-\theta_{\tau s}[F_{\tau L}^{j\dagger}(\hat{p},\alpha)e^{-i\alpha\cdot\hat{x}_{\tau L}}]\,[e^{i\alpha\cdot\hat{x}_{s L}}F_{s L}^{k}(\hat{p},\alpha)]\nonumber\\
&-\theta_{s \tau}[F_{s R}^{k}(\hat{p},\alpha)e^{i\alpha\cdot\hat{x}_{s R}}]\,[e^{-i\alpha\cdot\hat{x}_{\tau R}}F_{\tau R}^{k\dagger}(\hat{p},\alpha)]\Big)\bigg\},
\label{eq:translcov}
\end{align}
where
\begin{align}
F^k_{\tau \mu}(\hat{p},\alpha)=\int d\beta\,\tilde{\mathcal{A}}_{\tau}^{k}(\alpha,\beta)e^{i\beta \cdot \hat{p}_{\mu}}\label{eq:Fkmu}
\end{align}
is a completely general operator valued function of the operator $\hat{p}$.
Equation~\eqref{eq:translcov} fully characterizes translation covariant CP Gaussian maps.

\prlsec{Boost-covariance} 
Restricting to the subgroup of boosts $ \mathcal{B}\subset \mathcal{G}$ we have that: 
\begin{align}
\mathcal{U}_{\vec{b}}[\hat{x}_{t}] & =\hat{x}_{t}+t\,\vec{b}/{m},\\
\mathcal{U}_{\vec{b}}[\hat{p}] & =\hat{p}+\vec{b}.
\end{align}
where $\vec{b}=m\vec{v}$ is a momentum vector (a particle of mass $m$ boosted with velocity $\vec{v}$) and $\mathcal{U}_{\vec{b}}$ denotes the corresponding linear transformation (see Eq.~\eqref{eq:semi}). Imposing boost covariance, and following the analogous steps as for the characterization of translational covariance, we obtain the following equality: 
\begin{align}
\mathcal{D}(\alpha_{1},\alpha_{2},\beta_{1},\beta_{2},\tau,s)=
&\delta(\beta_{1}-\beta_{2}+\alpha_1 \frac{s}{m}-\alpha_2 \frac{\tau}{m}) \nonumber\\
& \mathcal{D}_{B}(\alpha_{1},\alpha_{2},\beta_{1},\beta_{2},\tau,s)\label{eq:Dfcond}
\end{align}
where $\mathcal{D}_{B}$ is a complex valued function. Performing the following change of variables: $\vec{\beta}_{1} \to \vec{\beta}_{1}-\tau{\vec{\alpha}_{1}}/{m}$ and $\vec{\beta}_{2} \to \vec{\beta}_{1}-s{\vec{\alpha}_{2}}/{m}$, using Eq.~\eqref{eq:Dfcond}, we can then rewrite Eq.~\eqref{eq:rho_formal2} as: 
\begin{align}
\mathcal{M}_{t}= & \exp_{+}\bigg\{\int_{0}^{t}d\tau\int_{0}^{t}ds\int_{\mathbb{R}^3} d\beta\,D_{jk}(\tau,s)\nonumber\\
&\Big([F_{s L}^{k}(\hat{x}_{s},\beta)e^{i\beta\cdot\hat{p}_{L}}][F_{\tau R}^{j\dagger}(\hat{x}_{\tau},\beta)e^{-i\beta\cdot\hat{p}_{R}}]\nonumber \\
 & -\theta_{\tau s}[F_{\tau L}^{j\dagger}(\hat{x}_{\tau},\beta)e^{-i\beta\cdot\hat{p}_{L}}]\,[e^{i\beta\cdot\hat{p}_{L}}F_{s L}^{k}(\hat{x}_{s},\beta)]\nonumber\\
&-\theta_{s \tau}[F_{s R}^{k}(\hat{x}_{s},\beta)e^{i\beta\cdot\hat{p}_{R}}]\,[e^{-i\beta\cdot\hat{p}_{R}}\,F_{\tau R}^{k\dagger}(\hat{x}_{\tau},\beta)]\Big)\bigg\},
\end{align}
where 
\begin{equation}
F^k_{\tau \mu}(\hat{x}_{\tau},\beta)=\int d\alpha\,\tilde{\mathcal{A}}_{\tau}^{k}(\alpha,\beta-\tau \alpha/m)e^{i\alpha \cdot \hat{x}_{\tau \mu}}
\label{eq:Fkmu2}
\end{equation}
is a completely general operator valued function of the operator $\hat{x}_{\tau}$. This equation completely characterizes boost covariant CP Gaussian maps.

\prlsec{Translation-boost Covariance}
We now require both translation and boost covariance. The dynamical
map $\mathcal{M}_{t}$ must satisfy condition Eq.~\eqref{eq:condexpt}
as well as condition Eq.~\eqref{eq:Dfcond}, \textit{i.e.},
\begin{align}
\mathcal{D}(\alpha_{1},\alpha_{2},\beta_{1},\beta_{2},\tau,s)=
&\delta(\alpha_{1}-\alpha_{2})\nonumber\\
&\delta(\beta_{1}-\beta_{2}+\alpha_1 \frac{s}{m}-\alpha_2 \frac{\tau}{m}) \nonumber\\
&\mathcal{D}_{TB}(\alpha_{1},\alpha_{2},\beta_{1},\beta_{2},\tau,s)\label{eq:boost-trasl}
\end{align}
Replacing Eq.~\eqref{eq:boost-trasl} in Eq.~\eqref{eq:rho_formal2}, performing again the following change of variables: $\vec{\beta}_{1} \to \vec{\beta}_{1}-\tau{\vec{\alpha}_{1}}/{m}$ and $\vec{\beta}_{2} \to \vec{\beta}_{1}-s{\vec{\alpha}_{2}}/{m}$, one obtains 
\begin{align}
\mathcal{M}{}_{t}= & \exp_{+}\Big\{\int_{0}^{t}d\tau\int_{0}^{t}ds\int d\alpha\int d\beta\,\mathcal{F}(\alpha,\beta,\tau,s)\nonumber\\
&\hspace{1cm}\Big(e^{i(\alpha\cdot\hat{x}_{s L}+\beta\cdot\hat{p}_{L})}e^{-i(\alpha\cdot\hat{x}_{\tau R}+\beta\cdot\hat{p}_{R})}\nonumber \\
 &\hspace{1cm}-\theta_{\tau s}e^{-i(\alpha\cdot\hat{x}_{\tau L}+\beta\cdot\hat{p}_{L})}e^{i(\alpha\cdot\hat{x}_{s L}+\beta\cdot\hat{p}_{L})}\nonumber\\
&\hspace{1cm}-\theta_{s \tau}e^{-i(\alpha\cdot\hat{x}_{s R}+\beta\cdot\hat{p}_{R})}e^{i(\alpha\cdot\hat{x}_{\tau R}+\beta\cdot\hat{p}_{R})}\Big)\bigg\},\label{eq:rho_formal3}
\end{align}
where $\mathcal{F}(\alpha,\beta,\tau,s)$ is a completely general real valued function. In this case the functional dependence of the map on the position and momentum operator is fixed~\footnote{Clearly, Eq.~\eqref{eq:rho_formal3} can be written in many different but equivalent ways by using unitary transformation on the operators.}.
This equation fully characterizes translation and boost covariant
CP Gaussian maps.

\prlsec{Rotation Covariance}
For completeness, we also discuss rotation covariance. Restricting to the subgroup of rotations $\mathcal{R}\subset\mathcal{G}$ we have:
\begin{align}
\mathcal{U}_{R}[\hat{x}_{s}] & =R \hat{x}_{s},\label{eq:r1}\\
\mathcal{U}_{R}[\hat{p}] & =R\hat{p},\label{eq:r2}
\end{align}
where $R$ is a generic rotation matrix and $\mathcal{U}_{R}$ the corresponding linear transformation (see Eq.~\eqref{eq:semi}). Using the relation $\vec{a} \cdot (R \vec{b}) = (R^{-1}\vec{a}) \cdot  \vec{b}$, where $\vec{a}$, $\vec{b}$ are generic vectors, and recalling that the integral measure $d\alpha d\beta$ is invariant under rotations, we perform the change of variables $\alpha\to R\alpha$, $\beta\to R\beta$ in Eq.~\eqref{eq:1cond}, which gives the condition: 
\begin{align}
\mathcal{D}(R\alpha_{1},R\beta_{1},R\alpha_{2},R\beta_{2},\tau,s)=\mathcal{D}(\alpha_{1},\beta_{1},\alpha_{2},\beta_{2},\tau,s).
\label{eq:D_rotations}
\end{align}
Equation~\eqref{eq:rho_formal2}, with the function $\mathcal{D}$ satisfying the symmetry given by Eq.~\eqref{eq:D_rotations}, characterizes rotational covariant CP Gaussian maps. This concludes the characterization of CP Gaussian maps covariant under Galilean symmetries.

\prlsec{Markovian limits} 
The CP Gaussian covariant 
maps derived here above reduce to the well-known Markovian CP Gaussian covariant maps in the Markovian limit. In particular, we immediately re-obtain the Holevo structures for the generators of the covariant quantum dynamical semigroup by imposing the request of Markovianity as given by Eq.~\eqref{eq:markovianity}. Under this assumption it is straightforward to show that Eq.~\eqref{eq:translcov} reduces to $\mathcal{M}_{t}=\exp_{+}\{\int_{0}^{t} d s \!\!\ \mathcal{L}_{s}\}$, where
\begin{align}
\mathcal{L}_{s}=&\int\!d\alpha\,\tilde{D}_{jk}(s)
\Big(F_{s L}^{k}(\hat{p},\alpha)e^{i\alpha\cdot\hat{x}_{s L}}F_{s R}^{j\dagger}(\hat{p},\alpha)e^{-i\alpha\cdot\hat{x}_{s R}}\nonumber\\
&-\frac{1}{2} F_{s L}^{j\dagger}(\hat{p},\alpha) F_{s L}^{k}(\hat{p},\alpha)
-\frac{1}{2} F_{s R}^{k}(\hat{p},\alpha) F_{s R}^{k\dagger}(\hat{p},\alpha)\Big),
\label{eq:holevot}
\end{align}
is the generator of the translational covariant semigroup.

Analogously, by considering the Markovian limit of the boost and translation covariant map derived in Eq.~\eqref{eq:rho_formal3}, we obtain the following generator:
\begin{align}
\mathcal{L}_{s}= &\int d\alpha\int d\beta\,\tilde{\mathcal{F}}(\alpha,\beta,s)\nonumber\\
&\Big(e^{i(\alpha\cdot\hat{x}_{s L}+\beta\cdot\hat{p}_{L})}e^{-i(\alpha\cdot\hat{x}_{s R}+\beta\cdot\hat{p}_{R})}-1\big),
\label{eq:holevotb}
\end{align}
where $\tilde{\mathcal{F}}$ is a positive valued function.
Equations.~\eqref{eq:holevot}, and~\eqref{eq:holevotb}
correspond to the Holevo results for covariance under translation and boost-translation, respectively~\cite{holevo1995translation,holevo1996covariant}.

\prlsec{Macroscopicity measure} More and more experiments are nowadays probing quantum mechanics in novel regimes, exploring in particular the boundary between quantum and classical~\cite{curceanu2015x,torovs2017bounds,carlesso2016experimental,bilardello2016bounds,belli2016entangling}. 
It becomes relevant to define a \emph{measure} that quantifies how far a given experiment pushes this boundary. This is a nontrivial task: what is the measure of \emph{macroscopicity} that correctly accounts for complexity, size, mass or some other feature of the system being explored?

Beginning with Leggett~\cite{leggett1980macroscopic,leggett2002testing} several measures of macroscopicity have been proposed~\cite{dur2002effective,bjork2004size,korsbakken2007measurement, marquardt2008measuring,lee2011quantification,nimmrichter2013macroscopicity,nimmrichter2014macroscopic}.
Among them, the one given by Nimmrichter and Hornberger in Ref.~\cite{nimmrichter2014macroscopic} has become quite popular in the matter-wave interferometry community because of its simplicity and versatility: they define as a \emph{macroscopicity measure} a real number that quantifies how well an experiment tests a \emph{minimal modification of quantum mechanics}. Specifically, they suggest the following measure:
\begin{align}\label{eq:macroscopicity}
\mu= \log(\tau/1\,s)
\end{align}
with $\tau$ the biggest excluded \emph{time scale} in which quantum superpositions are suppressed by the {minimal modification of quantum mechanics}.

They further assume that the \emph{minimal modification of quantum mechanics}, for a single particle with mass $m$, is described by a Markovian non-unitary TP, CP, Galilean covariant (translations, boosts and rotations) and time translation invariant map. This amounts to the non-unitary map generated by Eq.~\eqref{eq:holevotb}, where they choose the following parametrization of the correlation function:
\begin{equation}\label{eq:gtau}
\tilde{\mathcal{F}}(\alpha,\beta,s)=\frac{1}{\tau} g(\alpha,\beta),
\end{equation}
where $g$ is a positive, isotropic phase-space distribution normalized to unity (a Gaussian function with variances $\sigma_{\alpha}$, $\sigma_{\beta}$) and $\tau$ gives the \emph{time scale} in which superpositions are suppressed by the {minimal modification} (for further details see Refs.~\cite{nimmrichter2013macroscopicity,nimmrichter2014macroscopic}).

The measure $\mu$ defined in Eq.~\eqref{eq:macroscopicity} thus relies on the assumptions characterizing the {minimal modification}. Among these, Markovianity and Galilei covariance are usually taken for granted as they are a building block of the most successful non-relativistic theories: quantum and classical mechanics.
However, technological advances have come to the point of questioning the validity of these two assumptions; on top of this, minimal modifications need not satisfy them a priori. We take an example from the literature of collapse models, which can be seen as instances of \emph{minimal modifications} of quantum mechanics in the spirit of~\cite{nimmrichter2013macroscopicity}.
X-ray measurements~\cite{PhysRevC.59.2108} pose rather strong bounds on the collapse parameters~\cite{curceanu2015x} however the strength of the bounds depends critically on whether the collapse model is Markovian or not~\cite{adler2009photon, donadi2014emission}. The reason is that such experiments explore the $\approx 10^{18}$Hz region of the spectrum, meaning that the time resolution which is probed is $\approx 10^{-18}s$. Any cutoff in the spectrum of the collapse noise smaller than such frequencies weakens significantly the bound. A similar behavior is expected to occur for a macroscopicity measure that correctly includes non-Markovian effects. 
Markovianity might be verified only under a suitable temporal coarse graining of the underlying dynamics.
In general if the time resolution of the experiment is longer than the correlation times associated to the modifications of the theory, then the Markovian assumption is justified, as any non-Markovian dynamics with finite correlation times may be approximated by its Markovian limit~\footnote{see Supplemental Material S1, for the discussion on the long time behavior of the statistical operator $\rho_{t}$ under the a generic Gaussian evolution in Eq.~\eqref{eq:df}}. 

The assumption of Galilean covariance (translation and boost), even if it seems an innocent assumption, forces the non-unitary dynamics to produce an infinite growth of the system's energy on long time scales~\footnote{see Supplemental Material S2, for a study on the asymptotic behavior of the energy for the Galilei-Covariant dynamic in Eq.~\eqref{eq:rho_formal2}}. 
Galilean covariant maps must be then understood only as a good approximation that can be used in experiments that run for sufficiently short times, such that dissipative phenomena are negligible.
In experiments with a long running time, the results could be influenced by dissipative phenomena and consequently the assumption of Galilean covariant dynamics is too restrictive.
We consider a second example taken from collapse models. A recent experiment succeeded to cool a cloud of cold atoms to temperature less than $50^{+50}_{-30}$ pK~\cite{kovachy2015matter}. They measured the spreading of the cloud over time, which would be affected by modification of quantum mechanics. The analysis performed in Ref.~\cite{bilardello2016bounds} shows that the predictions of collapse models depend on whether dissipative effects are taken into account (Fig.~(8) of Ref.\cite{bilardello2016bounds} shows that the bounds on the collapse model drastically change with the thermalization temperature $T$, which quantifies the dissipation in the model).
Again, a similar dependence on dissipation is expected by a macroscopicity measure, which takes dissipative effects into account.

To summarize, although $\mu$ is a reasonable choice for the measure of macroscopicity in many instances, novel experiments probing the very short and very long time scales need a different measure of macroscopicity due to non-Markovian and dissipative effects, respectively. For such cases we propose to use the (translational covariant and non-Markovian) map given in Eq.~\eqref{eq:translcov} as the \textit{minimal modification}, with an appropriately chosen correlation function $D(t,s)$ and operators $F_{\tau \mu}(\hat{p},\alpha)$, where for simplicity we consider that the sum over $j$, $k$ contains only one term. We can still use Eq.~\eqref{eq:macroscopicity} to define the measure of macroscopicity, where now $\tau \to \tau (\tau_{c},T)$ is the biggest excluded \emph{time scale}, for fixed parameters $(\tau_{c},T)$, in which quantum superposition are suppressed by the minimal modification.
Here $\tau_{c}$ is the correlation time of the correlation function $D(t,s)$ and $T$ is the temperature measuring dissipative effects.
 
To be more concrete we suggest the exponential correlation function
\begin{align}\label{eq:corrfe}
D(t,s)=\frac{1}{2\tau_c} e^{-|t-s|/\tau_c}
\end{align}
and  the Gaussian operators
\begin{align}\label{eq:JoperatorRe}
F_{\mu}(\hat{p},\alpha)=\sqrt{\frac{1}{{\tau}}\frac{m^{2}}{m_{0}^{2}}\left(\frac{r_{c}}{\sqrt{\pi}\hbar}\right)^{3}} e^{-\frac{r_{c}^{2}}{2\hbar^{2}}[(1+k_{T})\alpha+2k_{T}\hat{\bm{p}}_{\mu}]^{2}},
\end{align}
where $k_T=\frac{\hbar^{2}}{8 m_0 r_{c}^{2}k_{B} T}$, $m_0=1$ amu is a reference mass, $k_{B}$ is Boltzmann's constant, $r_{c}$ is a free length parameter analogous to the spread $\sigma_{\alpha}$ in Eq.~\eqref{eq:gtau} and $\tau$ gives the \emph{time scale} in which the superpositions of a reference object with mass m are suppressed.
In the Markovian ($\tau_{c}\to 0$) and non-dissipative  ($T\rightarrow \infty$) limit, we reobtain the measure of macroscopicity proposed by Nimmrichter and Hornberger with $\sigma_{\beta} \rightarrow 0$ (see Eqs.~\eqref{eq:macroscopicity},~\eqref{eq:gtau}). 
 
This new measure depends critically on the values of $\tau_c$ and $T$. To illustrate this, we have studied the classicalization map in the regime of small distances and low momentum transfer in one spatial dimension~\footnote{We Taylor expand the operators in the time ordered exponential up to quadratic order in position and momentum operators, \textit{i.e.} we keep only the terms proportional to $\oldhat{x}$, $\oldhat{p}$, $\oldhat{x}^2$, $\oldhat{x}\oldhat{p}$, $\oldhat{p}\oldhat{x}$, $\oldhat{p}^2$.}. Specifically, we have considered a simple ideal experiment capable of resolving the time evolution of the spread of the wave-packet of a freely evolving particle. The associated macroscopicity measure is investigated in the non-Markovian and dissipative regimes  (cf. Supplemental Material S4 and Fig. S2), showing how it depends on the correlation time $\tau_c$ and temperature $T$.

\prlsec{Summary}
We have analyzed Galilean symmetries in non-Markovian Gaussian CP maps. The two main results of this letter are the characterization of translational and of Galilei (translation-boost) covariant non-Markovian CP Gaussian maps given by Eqs.~\eqref{eq:translcov} and \eqref{eq:rho_formal3}, respectively. These maps are a generalization of the well known Holevo results, which we reobtain in the Markovian limit. We have also provided the corresponding unravelling given by stochastic Schr\"odinger equations in a form suitable for non-perturbative numerical analysis~\footnote{see Supplemental Material S3 for the study on the stochastic unraveling that leads to CP gaussian maps satisfying translation, boost, boost-translation or rotation symmetry} 
As mentioned in the introduction, these results can find applications in several fields of research \cite{vacchini2001translation, de2013derivation, yan2000unified, yan2000unified1, Sandulescu1987277,474887, vacchini2009quantum, 6b00498}. We have also analyzed the role that non-Markovian and dissipative effects play in the construction of a macroscopicity measure. We have shown that experiments probing the quantum-to-classical boundary on very short or very long time scales might not be adequately described by the macroscopicity measure in Ref.~\cite{nimmrichter2013macroscopicity}, and a more general definition is needed, as the one we propose, based on Eqs.~\eqref{eq:translcov},~\eqref{eq:corrfe}, and~\eqref{eq:JoperatorRe}.

\prlsec{Acknowledgements} The authors acknowledge A. Smirne, and A. Tilloy for insightful discussions. They also thank L. Ferialdi for insightful discussions  and for the \emph{Mathematica} codes of the {\scriptsize CD}QMUPL dynamics and gratefully acknowledge financial support from the University of Trieste (grant FRA 2016) and INFN. G.G. acknowledges financial support from ICTP Trieste. G.G. and M.T. contributed equally to this work.

\bibliographystyle{unsrt}
\bibliography{gcm}

\end{document}


\begin{center}
\textbf{\large Supplemental Material for\\ ``General Galilei covariant Gaussian maps''}
\end{center}

\setcounter{equation}{0}
\setcounter{figure}{0}
\setcounter{table}{0}
\setcounter{section}{0}
\setcounter{page}{1}
\makeatletter
\renewcommand{\thesection}{S\arabic{section}}
\renewcommand{\theequation}{S\arabic{equation}}
\renewcommand{\thefigure}{S\arabic{figure}}
\renewcommand{\bibnumfmt}[1]{[S#1]}
\renewcommand{\citenumfont}[1]{S#1}

\newcommand{\avs}[1]{\mathbb{E}_{\mathbb{Q}}\left[\mathcal{T}({#1})\right]} 
\newcommand{\abs}[1]{\left| {#1} \right|}

\section{Long Time Evolution and short correlation time}
\label{sec:Long-Time-Evolution}
We discuss the long time behavior of the statistical operator $\oldhat{\rho}_{t}$ under the a generic Gaussian evolution given by Eq.~(1), which we rewrite as
\begin{equation}
\mathcal{M}_{t}=\mathcal{T}\exp\left\{ \frac{1}{2} \int_{0}^{t}d\tau\int_{0}^{t}ds \, \mathcal{V}(\tau,s) \right\}, 
\label{eq:df2}
\end{equation}
where $\mathcal{T}$ denotes the time order operator and
\begin{equation}
\mathcal{V} (\tau,s)= D_{jk}(\tau,s)
\left(\oldhat{A}_{sL}^{k}\oldhat{A}_{\tau R}^{j}-\theta_{\tau s}\oldhat{A}_{\tau L}^{j}\oldhat{A}_{sL}^{k}-\theta_{s \tau}\oldhat{A}_{sR}^{k}\oldhat{A}_{\tau R}^{j}\right),
\end{equation}
under the assumption that there exists a finite (positive) time $\tau_c$ (the correlation time) such that:
\begin{equation}
\mathcal{V} (\tau,s) \approx \;0 \;\;\;\;\text{when}\;\; |\tau-s|>\tau_c. \label{eq:tauc}
\end{equation}
Specifically we study Eq.~\eqref{eq:df2} in the regime:
\begin{equation}
t \gg \;\tau_c .\label{eq:tggtauc}
\end{equation}
We argue that under these assumptions, the map in Eq.~\eqref{eq:df2} can be approximated up to $\mathcal{O}(\tau^{2}_{c}/t^{2})$ by the Markovian map generated by:
\begin{align}\label{eq:markpp}
\partial_{t}\hat{\rho}_{t}={\bf{L}}_{t}\hat{\rho}_{t}
\end{align}
where:
\begin{align}
{\bf{L}}_{t}=\mathcal{T}\left(\int_{0}^{\infty}d\tilde{\tau}\, \mathcal{V}(t,t-\tilde{\tau}) \right)
\end{align}
Here we sketch the derivation, leaving a rigorous analysis of the limits of validity of this result for future research. 

In order to show the validity of Eq.~\eqref{eq:markpp} we take the time derivative of $\hat{\rho}_{t}=\mathcal{M}_{t}\hat{\rho}$ with $\mathcal{M}_{t}$ in Eq.~\eqref{eq:df2}, \textit{i.e.}
\begin{align}\label{eq:n_master}
\partial_{t}\oldhat{\rho}_{t}=\mathcal{T}\left(
 \int_{0}^{t} d\tilde{\tau}
\mathcal{V} (t,\tilde{\tau}) \, 
e^{\frac{1}{2}\int_{0}^{t} d\tau \int_{0}^{t} ds \, \mathcal{V} (\tau,s) } \right)
\oldhat{\rho}. 
\end{align}
and rewrite it as:
\begin{equation} 
\partial_{t}{\oldhat{\rho}}_{t}=\mathcal{T}\left( \int_{0}^{t} d\tilde{\tau}
\mathcal{V} (t,\tilde{\tau}) \, 
e^{\frac{1}{2}\int_{\tilde{\tau}}^{t} d\tau \int_{\tilde{\tau}}^{t} ds \, \mathcal{V} (\tau,s)}\, 
e^{\int_{\tilde{\tau}}^{t} d\tau \int_{0}^{\tilde{\tau}} ds \, \mathcal{V} (\tau,s)}\,
e^{\frac{1}{2}\int_{0}^{\tilde{\tau}} d\tau \int_{0}^{\tilde{\tau}} ds \, \mathcal{V} (\tau,s)}
 \right)
\hat{\rho}. 
\label{eq:approx_stocheq}
\end{equation}
We notice that under the assumption in Eq.~\eqref{eq:tauc} the first factor on the r.h.s. of Eq.~\eqref{eq:approx_stocheq} forces the variable $\tau$ to satisfy
\begin{align}
t \ge \tilde{\tau} \ge (t-\tau_{c}).
\end{align}
Thus we can make the approximations
\begin{align}
&\int_0^t d\tilde{\tau} \, \mathcal{V}(t,\tilde{\tau}) 
\approx \int_{t-\tau_c}^{t}d\tilde{\tau} \, \mathcal{V}(t,\tilde{\tau}) \propto \tau_c\nonumber\\
&\int_{\tilde{\tau}}^t d\tau \int_{\tilde{\tau}}^t ds \, \mathcal{V}(\tau,s) 
\approx \int_{t-\tau_c}^t  d\tau \int_{t-\tau_c}^t   ds \, \mathcal{V}(\tau,s) 
\propto \tau_{c}^2,\nonumber\\
&\int_{\tilde{\tau}}^t d\tau \int_0^{\tilde{\tau}} ds \, \mathcal{V}(\tau,s) 
\approx \int_{t-\tau_{c}}^{t}  d\tau \int_{{t}-2\tau_c}^{t-\tau_{c}}  ds \, \mathcal{V}(\tau,s) \propto \tau_{c}^2, \nonumber\\
&\int_{0}^{\tilde\tau}d\tau\int_{0}^{\tilde\tau}ds \mathcal{V}(\tau,s)\approx \int_{0}^{t-\tau_{c}}d\tau\int_{0}^{t-\tau_{c}}ds \,\mathcal{V}(\tau,s)\propto \tau_{c}t
\end{align}
Exploiting these results one is allowed to approximate Eq.~\eqref{eq:approx_stocheq} as 
\begin{equation}
\partial_{t}\oldhat{\rho}_t=
\mathcal{T}\left( \int_{t-\tau_c}^{t} d\tilde{\tau}
\mathcal{V} (t,\tilde{\tau})\right) \oldhat{\rho}_{\tilde{\tau}} +\mathcal{O}((\tau_c/t)^2), 
\label{eq:timennlocmm}
\end{equation}
In the above equation $\hat{\rho}_{\tilde{\tau}}$ can be replaced with $\hat{\rho}_{t}$, because under the assumption \eqref{eq:tggtauc}, $\hat{\rho}_{t}=\hat{\rho}_{\tilde{\tau}}+\mathcal{O}((\tau_{c}/t)^{2})$.

Replacing $\hat{\rho}_{\tilde{\tau}}\to \hat{\rho}_{t}$ and performing the change of variables $\tilde{\tau}\to (t-\tilde{\tau})$ in Eq.~\eqref{eq:timennlocmm} we obtain:
\begin{equation}
\partial_{t}\oldhat{\rho}_t= \mathcal{T}\left(\int_{0}^{\tau_{c}} d\tilde{\tau}
\mathcal{V} (t,t-\tilde{\tau})  \right) \oldhat{\rho}_t + \mathcal{O}((\tau_c/t)^2).
\label{eq:timennlocmm2111}
\end{equation}
Using the assumption in Eq.~\eqref{eq:tauc} one may now replace the  upper limit of the integral with $\infty$ and eventually get:
\begin{equation}
\partial_{t} \oldhat{\rho}_t= \left[\mathcal{T}\left(\int_0^{\infty} d\tilde{\tau}
\mathcal{V} (t,\tilde{\tau})  \right)\right] \oldhat{\rho}_t + \mathcal{O}((\tau_c/t)^{2}),
\label{eq:timennlocmm2}
\end{equation}
which is the desired result (see Eq.~\eqref{eq:markpp}).
When $\tau_c/t \to 0 $, the superoperator $\mathcal{V}(t,s)$ can be assumed delta correlated in time (Markovian assumption):
\begin{align}\label{eq:markv}
\mathcal{V}(t,s) \simeq \delta (t-s)\mathcal{\tilde{V}}(t)\quad \text{with} \quad \mathcal{\tilde{V}}\equiv \int_{0}^{\infty}d\tilde{\tau}\mathcal{V}(t,\tilde{\tau}),
\end{align}
which corresponds to Eq.~(2) in the main text. 
More specifically, in this limit it is then straightforward to obtain from Eq.~\eqref{eq:timennlocmm2}:
\begin{equation}\label{eq:lindblad1}
\partial_{t} \oldhat{\rho}_t = \tilde{D}_{jk}(t)
\left(\oldhat{B}_{t}^{k}\oldhat{\rho}_t \oldhat{B}_{t}^{j}-
\frac{1}{2}\{\oldhat{B}_{t}^{k}\oldhat{B}_{t}^{j},\oldhat{\rho}_t\}\right).
\end{equation}
where $\oldhat{B}^{k}_{t}$ and $\tilde{D}_{jk}(t)$ are implicitly defined through  Eq.~\eqref{eq:markv}.
Further assuming that the generator of the dynamics does not depend on time, \textit{i.e.} $\tilde{D}_{jk}(t)=\tilde{D}_{jk}$,  we obtain the standard Lindblad form of the generator. 

We stress that this procedure does not guarantee complete positivity of the approximated map in the long time scale given by  Eq.~\eqref{eq:lindblad1}. This reflects the fact that the approximated dynamics is valid only for states $\oldhat{\rho}_{t}$ that are the image of the full dynamical map $\mathcal{M}_{t}$, with $t\gg\tau_{c}$, of the initial state $\hat{\rho}_0$ at time $t=0$. A similar analysis has already been done by Di\'{o}si~\cite{diosi1993high} on the Caldeira Legget master equation in the attempt to extend the original result~\cite{caldeira1983path} to a medium temperature environment. In other words, loosely speaking, the non-Markovian character of the evolution gets encoded over time in the state $\oldhat{\rho}_t$ as well as in the matrix $\tilde{D}_{j k}$ and operators $\oldhat{B}_t^k$ of the approximate Markovian dynamics given by  Eq.~\eqref{eq:lindblad1}.

\section{Dissipation and Covariance}
\label{sec:dissip_cov}
We study the asymptotic behavior of the energy under the assumption of Galilei-Covariance and finite correlation time.
Under the assumption of finite correlation time $\tau_{c}$, the dynamical map can be treated as Markovian in the long time limit, as is shown in Sec.~\ref{sec:Long-Time-Evolution}. This allows to investigate the asymptotic behavior of the energy by considering Markovian dynamical maps, to which Holevo's characterization of Markovian master equations can be applied \cite{holevo1996covariant,holevo1995translation,holevo1993conservativity, holevo1993note}. Specifically, we consider a Galilei covariant Markovian master equation (in the Schr\"{o}dinger picture):
\begin{equation}
\frac{d}{dt}\hat{\rho}_{t}=-\frac{i}{\hbar}[\oldhat{H},\hat{\rho}_{t}] 
+ \!\int\!d\oldalpha\!\int\!d\oldbeta\mathcal{F}(\oldalpha,\oldbeta)\Big(e^{i(\oldalpha\oldhat{x}+\oldbeta\oldhat{p})} \hat{\rho}_{t}e^{-i(\oldalpha\oldhat{x}+\oldbeta\oldhat{p})}-\hat{\rho}_{t}\Big).\label{eq:holevotb2}
\end{equation}

For simplicity we limit to the 1-D case (the 3-D case is a straightforward generalization) of an isolated free system: specifically, we set $\oldhat{H}=\frac{\oldhat{p}^{2}}{2m}$. It is then easy to obtain the following system of equations for the expectation values $\braket{\oldhat{p}^{2}}_{t}\equiv \text{Tr}[\oldhat{p}^{2}\hat{\rho}_{t}]$ and $\braket{\oldhat{p}}_{t}\equiv \text{Tr}[\oldhat{p}\hat{\rho}_{t}]$:
\begin{align}
\partial_{t}\braket{\oldhat{p}^{2}}_{t}&= 2\lambda \braket{\oldhat{p}}_{t}+ \gamma,\nonumber\\
\partial_{t}\braket{\oldhat{p}}_{t}&= \lambda,
\end{align}
where we have defined $\lambda= \int d\oldalpha d\oldbeta\, \oldalpha\, \mathcal{F}(\oldalpha,\oldbeta)$ and $\gamma= \int d\oldalpha\int d\oldbeta\, \oldalpha^{2}\,\mathcal{F}(\oldalpha,\oldbeta)$. Solving the above system of equations we obtain the following evolution for the energy:
\begin{align}
\braket{\oldhat{p}_{t}^{2}}&= \lambda^{2}t^{2}+(\lambda \braket{\oldhat{p}}_{0}+\gamma) t + \braket{\oldhat{p}^{2}}_{0}.
\end{align}
This equation diverges for $t\to \infty$, showing that Galilei-Covariance leads to an unbounded growth of the energy for long times. 

It is then necessary to relax one of the hypothesis in order to have a physically consistent behavior for long time scales. However, the translational covariance of the model is needed for the reproducibility of the experiments, leaving boost covariance as the only assumption that can be relaxed. For example, this has been done in ~\cite{smirne2015dissipative}, where boost covariance is explicitly broken.

\section{Stochastic Schr\"odinger equations}
We characterize the stochastic unraveling that leads to CP gaussian maps satisfying  translation, boost, boost-translation or rotation symmetry.
We notice that a CP Gaussian map, covariant under Galilean symmetries, can always be written in the form of Eq.~(4), with the matrix $D$ implementing the symmetries: for translation, boost, boost-translation or rotation symmetry the function $D$ has to satisfy Eqs.~(15), (21), (24), or (28), respectively.
Following~\cite{diosi2014general}, and exploiting Eq.~(4) it is straightforward to write the unraveling: 
\begin{align}
i\partial_{t}\ket{\psi_{\xi}(t)}= &\int d\Gamma_{1}
e^{i(\alpha_{1}\cdot\hat{x}_{t}+\beta_{1}\cdot\hat{p})} 
\times\bigg[\xi_{t}(\Gamma_{1}) 
+ \int\!\!d\Gamma_{2}\!\int\!\!ds\, \Big(D(\Gamma_{1},\Gamma_{2},t,s)
-S(\Gamma_{1},\Gamma_{2},t,s)\Big)
\frac{\olddelta}{\olddelta\xi_{s}(\Gamma_{2})}\bigg]\ket{\psi_{\xi}(t)},
\label{eq:psi}
\end{align}
where $\Gamma_{j}$ denotes collectively the two vectors $\alpha_{j}$, $\beta_{j}$ and $\xi_{t}(\Gamma_{j})$ is a complex valued Gaussian stochastic field with zero mean and variance:
\begin{align}
\mathbb{E}_{\xi}[\xi_{t}(\Gamma_{1})\xi_{s}(\Gamma_{2})]=&S(\Gamma_{1},\Gamma_{2},t,s),\nonumber\\
\mathbb{E}_{\xi}[\xi^{*}_{t}(\Gamma_{1})\xi_{s}(\Gamma_{2})]=&D(\Gamma_{1},\Gamma_{2},t,s)
\end{align}
where $D$ characterizes the symmetries of the gaussian maps according to Eqs.~(15), (21), (24), or (28) and $S$ is constrained only by the request of positivity of the covariance matrix of the noise field $\xi$. 
However, Eq.~\eqref{eq:psi}, due to the presence of the functional derivative $\olddelta/\olddelta\xi_{s}$, cannot be trivially used for numerical analysis. 
To solve this problem, as discussed in \cite{tilloy2017markovian}, one can introduce an auxiliary stochastic field. Specifically, Eq.~\eqref{eq:psi} is equivalent to:  
\begin{align}\label{eq:phi}
i\partial_{t}\ket{\psi_{\xi,\eta}(t)}= \int d\Gamma_{1}
e^{i(\alpha_{1}\cdot\hat{x}_{t}+\beta_{1}\cdot\hat{p})}\bigg[\xi_{t}(\Gamma_{1})
+ \eta_{t}(\Gamma_{1})\bigg]\ket{\psi_{\xi,\eta}(t)},
\end{align}
where $\eta_{t}$ is a complex stochastic noise with zero mean and variance:
\begin{align}
\mathbb{E}_{\eta}[\eta_{t}(\Gamma_{1})\eta_{s}(\Gamma_{2})]&=K(\Gamma_{1},\Gamma_{2},t,s)+K(\Gamma_{2},\Gamma_{1},s,t),\nonumber\\
\mathbb{E}_{\eta}[\eta^{*}_{t}(\Gamma_{1})\eta_{s}(\Gamma_{2})]&=Q(\Gamma_{1},\Gamma_{2},t,s),
\end{align}
with:
\begin{align}
K(\Gamma_{1},\Gamma_{2},t,s)=\theta_{t s}[D(\Gamma_{1},\Gamma_{2},t,s)-S(\Gamma_{1},\Gamma_{2},t,s)],
\end{align}
while $Q$ is an arbitrary function. In particular, the unraveling given by Eq.~\eqref{eq:psi} can be generated from the unraveling given in Eq.~\eqref{eq:phi} by averaging over the auxilary noise field, \textit{i.e.}:
\begin{equation}
\ket{\psi_{\xi}(t)}=\mathbb{E}_{\eta}[\ket{\psi_{\xi,\eta}(t)}].
\end{equation}

\section{Measure of macroscopicity: colored and dissipative QMUPL}
Some of the most well-known classicalization maps are given by spontaneous collapse models~\cite{bassi2013models}. Among these models the Quantum-Mechanics-with-Universal-Position-Localization (QMUPL)  model~\cite{PhysRevA.40.1165,PhysRevA.42.5086} offers the possibility of obtaining simple analytical results, whilst retaining most of the important physical features of the more refined models. Specifically, we consider its colored and dissipative generalization (cdQMUPL) in one spatial dimension~\cite{PhysRevA.86.022108} and illustrate the necessity of studying the quantum-to-classical boundary in the the parameter space $(T, \tau_c)$ (see main text).  
  The cdQMUPL model dynamics for the statistical operator for a particle of mass $m$ is given by Eq.~(1), with the $j$,$k$ sum containing only one term and with the following replacements:
\begin{align}
\oldhat{A} \rightarrow\; &\oldhat{x}+\frac{i\epsilon}{\hbar}\oldhat{p} \label{eq:Aqmupl}
\\
\oldhat{H} \rightarrow\; &\oldhat{H}+ \frac{\lambda \epsilon}{2} \{\oldhat{x},\oldhat{p}\}  \label{eq:Hqmupl}
\\
D(\tau,s) \rightarrow\; &\lambda D(\tau-s) \label{eq:corrqmupl}
\end{align} 
The parameters $\lambda$, $\epsilon$ scale with the mass of the system $m$ as $\lambda=\frac{m_0}{m}\lambda_0$ and $\epsilon=\frac{m_0}{m}\epsilon_0$, respectively, where $m_0=1$ amu is a reference mass.
The parameter $\epsilon_0$ can be converted into a temperature $T= \frac{\hbar^2}{4 m_0 k_B \epsilon_0}$, where $k_B$ is Boltzmann's constant.
Moreover, we assume that the $D$ function in Eq.~\eqref{eq:corrqmupl} is the exponential correlation function in Eq.~(33) of the main text. 

This model can be alternatively understood as the a limiting case of small distances and low momentum transfer of the classicalization map defined by Eq.~(17) through Eqs.~(33) and (34). Specifically, the relation between the classicalization map proposed in the main text and cdQMUPL model's parameters is the following:  $\lambda=\gamma/(2r_{c}^{2} )$ and $\epsilon=2 r_{c}^{2} k_{T}$.
\begin{figure}[b]
\centering
\begin{minipage}[t]{0.5\columnwidth}
	\includegraphics[width=1.0\linewidth]{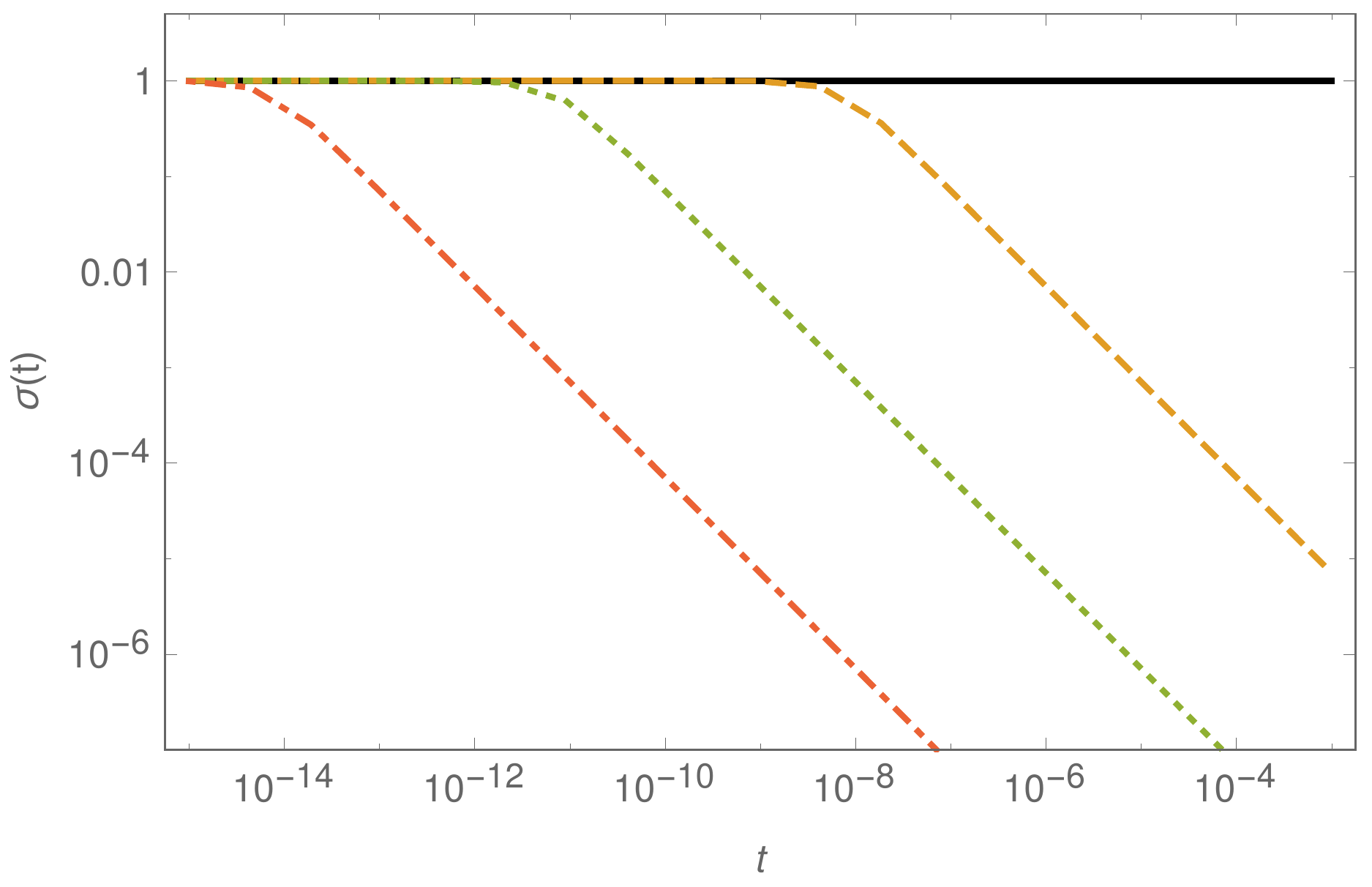}
\end{minipage}%
\begin{minipage}[t]{0.5\columnwidth}
	\centering
  \includegraphics[width=1.0\linewidth]{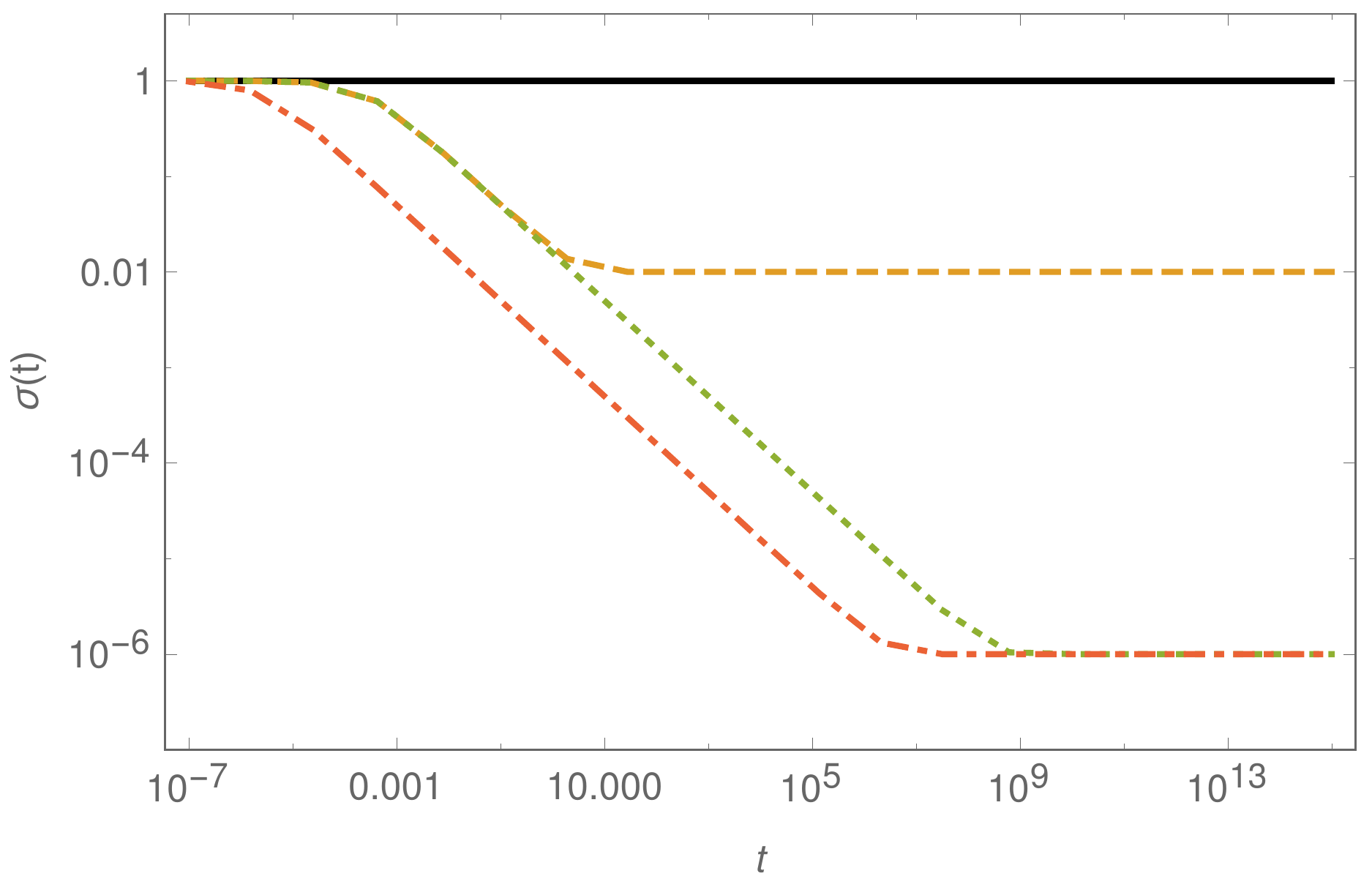}
\end{minipage}
\caption{
The evolution of the spread $\sigma(t)$ of a Gaussian wave-packet under the cdQMUPL classicalization map. The black solid line denotes the reference quantum mechanical evolution spread.
Left: We have set $\lambda= 10^{22} \text{m}^{-2}\text{s}^{-1}$ and $T \rightarrow \infty$ (non-Markovian regime). The orange dashed, green dotted and red dot-dashed lines denote the evolution for $\tau_c=10^{6}$ s, $\tau_c=1$ s and $\tau_c=10^{-6}$ s, respectively. Right: We have set  $\tau_c \rightarrow 0$ (dissipative regime). The orange dashed, green dotted and red dot-dashed denote the evolution for $(\lambda=10^{3} \text{m}^{-2}\text{s}^{-1}, T=10^{-7} \text{K})$, $(\lambda=10^{3} \text{m}^{-2}\text{s}^{-1}, T=10^{-15} \text{K})$ and $(\lambda=10^{5} \text{m}^{-2}\text{s}^{-1}, T=10^{-15} \text{K})$, respectively.}
\label{Fig:1}
\end{figure}
\begin{figure}[!t]
\centering
\begin{minipage}[t]{0.5\columnwidth}
	\includegraphics[width=0.5\linewidth]{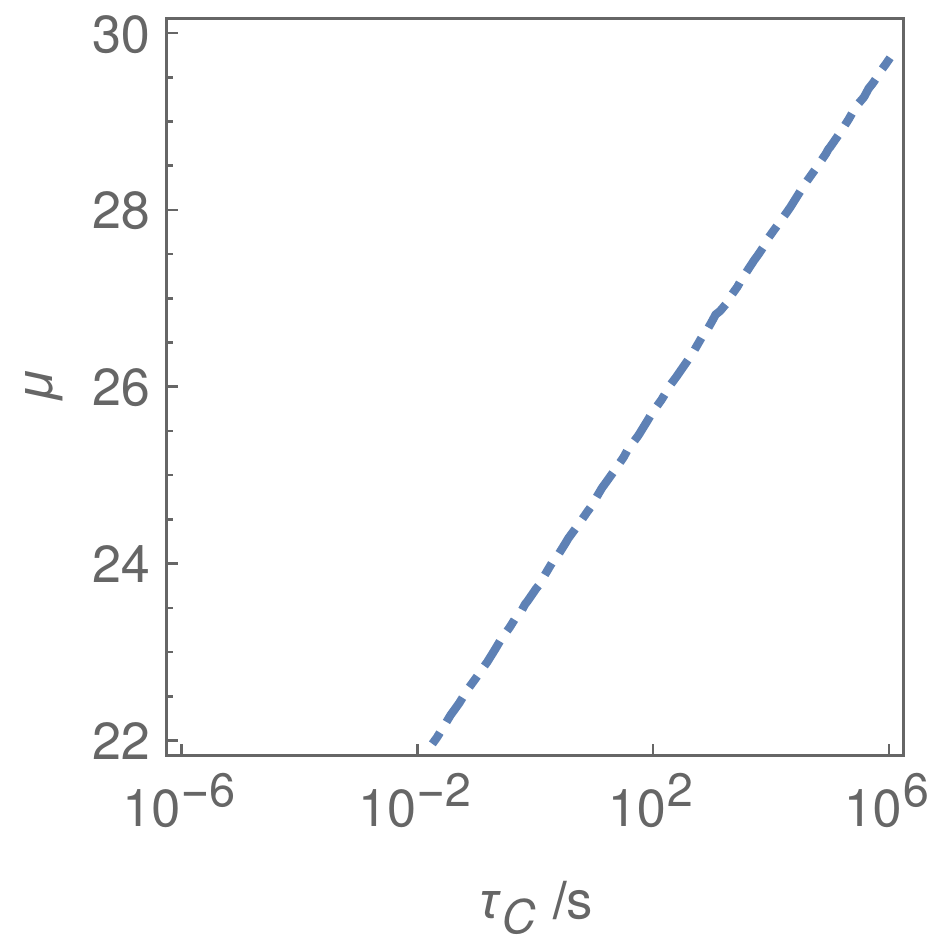}
\end{minipage}%
\begin{minipage}[t]{0.5\columnwidth}
	\centering
  \includegraphics[width=0.5\linewidth]{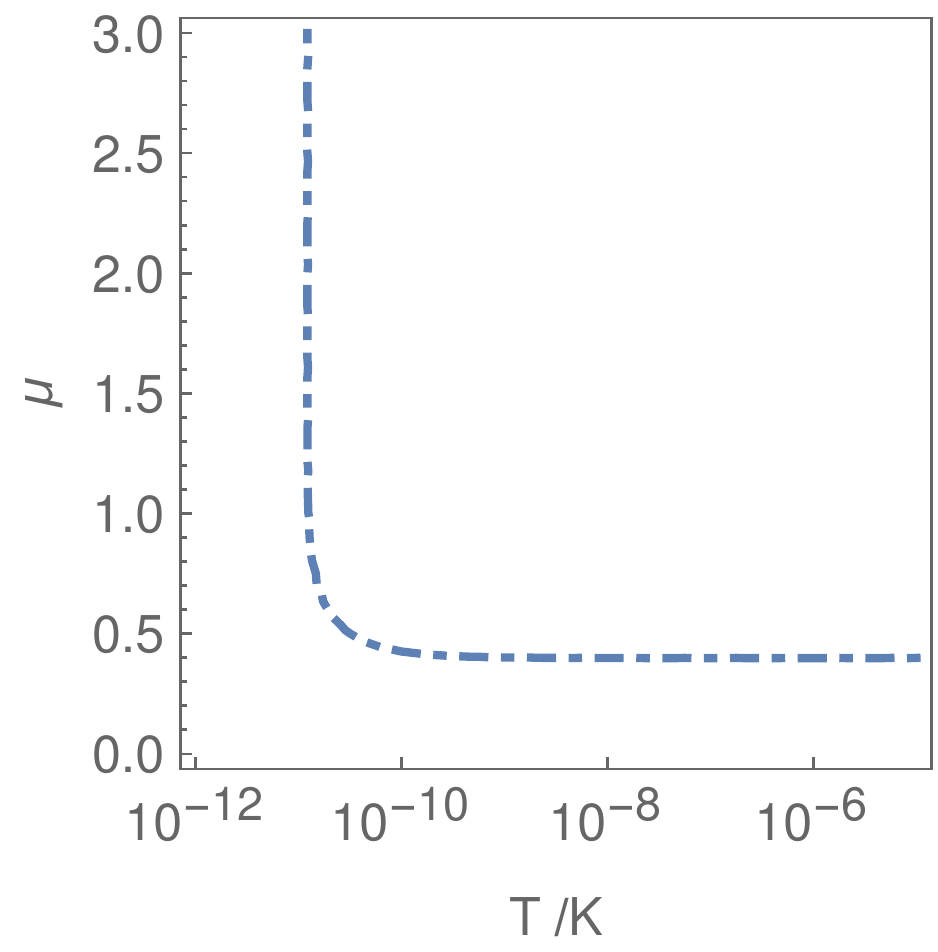}
\end{minipage}
\caption{We consider an experiment that can distinguish between the quantum mechanical spread $\sigma^{\text{(QM)}}_t$ and the cdQMUPL spread $\sigma^{\text{(cd)}}_t$, when $\sigma^{\text{(cd)}}_t/ \sigma^{\text{(QM)}}_t = 10^{-2}$. We plot the dependence of the cdQMUPL measure of macroscopicty (see main text Eqs.~(17), (31), (33) and (34)) with respect to the correlation time $\tau_c$ and temperature $T$. Left: We set $T\ rightarrow \infty $ and consider an experiment which lasts $t=10^{-10}s$ (non-Markovian regime). We see that the longer the correlation time the bigger the macroscopicity of the experiment.
Right: We set $\tau_c\rightarrow 0$ and consider an ideal experiment lasting $t=10^{3}s$ (near the dissipative regime). On the one hand, we see that for temperatures above $T=10^{-10}K$ we obtain a constant macroscopicty measure: this corresponds to the non-dissipative regime. On the other hand, below $T=10^{-10}K$ the situation changes drastically: we are in the dissipative regime, where the macroscopicity measure $\mu$ heavily depends on temperature $T$.}
\label{Fig:2}
\end{figure}

  We further notice that this model reduces to the known non-dissipative~\cite{bassi2009non} and Markovian~\cite{bassi2005energy} limits by imposing $T\rightarrow \infty $ and $\tau_{c}\to 0$, respectively. By imposing both conditions we reobtain the Markovian and non-dissipative QMUPL model.

We will now illustrate through this model that an adequate analysis on how well a given experiment explores the quantum-to-classical boundary must be performed in the parameter space $(\tau_{c}, T)$.

To this end we consider Gaussian wave-packets: the Gaussian form of the wave packet is preserved by the cdQMUPL evolution and the spread of the wave-packet $\sigma_t$ can be calculated analytically~\cite{PhysRevA.86.022108}. For the sake of the discussion, we consider an initial wave-packet with $\sigma_0=1$ m and look at the evolution in regimes, where non-Markovian and dissipative effects become relevant. 
From Fig.~\ref{Fig:1}, we see that, on the one hand, for small times (left) the evolution of the spread depends on the correlation time $\tau_{c}$, while, on the other hand, for large times (right) the evolution depends on the temperature $T$. We further illustrate this point in Fig.~\ref{Fig:2}, where we depict the behaviour of the measure of macroscopicty as a function of the correlation time $\tau_{c}$ and temperature $T$ in the non-Markovian and dissipative regimes, respectively, for an ideal measurement capable of resolving the spread of the wave function. As the figure shows, the measure heavily depends on $\tau_{c}$ and $T$, in situations where non-Markovian and dissipative effects dominate, respectively.

\bibliographystyle{unsrt}
\bibliography{gcm}